\documentclass[12pt]{iopart}

\usepackage{graphicx}

\def\pp#1#2{\left\leftdelim#1#2 \right\rightdelim#1}

\def\leftdelim#1{\ifcase #1.\or(\or
  [\or\{\or<\or\langle\or|\or\|\fi}    
\def\rightdelim#1{\ifcase #1.\or)\or
  ]\or\}\or>\or\rangle\or|\or\|\fi}    

\let\f\frac            
\def\F#1#2{#1/#2}   
\let\s\sqrt
\def\e#1{{\mathchoice{\hbox{\sl e\/}^{#1}}{\hbox    
  {\sl e\/}^{#1}}{\hbox{\scriptsize\sl e\/}^{#1}}   
  {\hbox{\tiny \sl e\/}^{#1}}}}                     
\def\ee{{\mathchoice{\hbox{\sl e\/}}{\hbox{\sl e\/}}    
  {\hbox{\scriptsize \sl e\/}}{\hbox{\tiny \sl e\/}}}}
\def\Ordo#1{{\cal O} \hskip -.15em \left( #1 \right)}
\def\d{{\rm d}}  

\def\cE{{\cal E}}

\def\h{\hbar}

\def\qH{{\rm H}}
\def\qplus{_+}
\def\qD{^{\rm (Dir)}}
\def\qN{^{\rm (Neu)}}

\def\qa{\tilde{\alpha}}
\def\qx{x}  \def\qX#1{#1_*}  \def\qxx{\qX{\qx}}
\def\qf{g}  \def\qp{p}

\begin{document}
\title{Quantum Force Induced on a  Partition Wall in a
Harmonic Potential}

\author{T F\"ul\"op$^1$
and I Tsutsui$^2$}

\vspace{5mm}

\address{$^1$ Montavid Thermodynamic Research Group, Igm\'andi
u.~26.~fsz.~4, 1112 Budapest, Hungary}
\ead{tamas.fulop@gmail.com} 

\vspace{5mm}

\address{$^2$ Institute of Particle and Nuclear Studies, High Energy
Accelerator Research Organization (KEK), Tsukuba 305-0801, Japan} 
\ead{izumi.tsutsui@kek.jp} 

\date{\today}

\begin{abstract}
Boundary effects in quantum mechanics are examined by considering a partition wall inserted at the centre of a harmonic oscillator system.   We put an equal number of particles on both sides of the impenetrable wall keeping the system under finite temperatures.  When the wall admits distinct boundary conditions on the two sides, then a net force is induced on the wall.  We study the temperature behaviour of the induced force both analytically and numerically under the combination of the Dirichlet and the Neumann conditions, and determine its scaling property for two statistical cases of the particles: fermions and bosons. We find that the force has a nonvanishing limit at zero temperature $T = 0$ and exhibits scalings characteristic to the statistics of the particles.  We also see that for higher temperatures the force decreases according to $1/\sqrt{T}$, in sharp contrast to the case of the infinite potential well where it diverges according to $\sqrt{T}$.  The results suggest that, if such a nontrivial partition wall can be realized, it may be used as a probe to examine the profile of the potentials and the statistics of the particles involved.
\end{abstract}


\maketitle

\section{Introduction}
\label{s1}

Quantum systems are often delineated by modeling their classical counterparts --- in fact, it is a standard practice that we define a system in quantum mechanics through the procedure called \lq quantization\rq, which amounts to replacing functions of phase space in classical mechanics by appropriate operators based on commutation relations.   However, 
this quantization procedure does not necessarily provide a unique quantum system to a given 
classical system, with the familiar example being the ordering ambiguity of operators.  Nontrivial topology of the classical configuration space furnishes an additional ambiguity in the  quantum system, which is exemplified by a particle moving on a circle where a multiple of windings are allowed for transitions.  In particle physics, the same topological effect is known to be responsible for the infinite vacua structure which causes the strong CP violation (see, {\it e.g.}, \cite{Jackiw}).    

Another source of ambiguity, which is less recognized and yet physically more tangible than the aforementioned ones, lies in the choice of boundary conditions imposed on quantum states \cite{RS,AG,AGHH}.    A prime example of this may be found in a particle system with an impenetrable wall, where one has a variety of boundary conditions to choose, say, from Dirichlet to Neumann or anything in between.  These boundary conditions represent different physical properties of the wall which are missing in the classical description, and can lead to novel effects in the quantum system \cite{FCT}.   A further example is provided by a singular point on a line, which serves as a source for a number of interesting properties, including supersymmetry and Berry phase \cite{CFT,Moebius,FT}.  
The importance of boundary conditions in quantum mechanics becomes evident if we recall the 
rudimentary fact that a different choice of boundary conditions yields a different spectrum.  The fact that the physical properties of quantum dots depend heavily on the boundary conditions  will also be worth mentioning.  

In order to provide a simple setup where the boundary effects can be observed directly, in our previous work  \cite{FTjpa}
we presented a case study of the induced pressure, or statistical quantum force, which emerges on an impenetrable partition wall inserted at the centre of an infinite potential well, when the wall is assumed to realize a nontrivial set of boundary conditions: the Dirichlet condition on one side and the Neumann on the other.   
We will not delve into here how such a partition can actually be manufactured as a device, but only mention that walls admitting generic boundary conditions including Dirichlet and Neumann can be realized by a combination of square well potentials in the vanishing limit of their widths \cite{FCT}.   When two such walls, one with Dirichlet and the other with Neumann, are \lq glued together\rq\ within a narrow distance, then it will serve as our partition effectively under a scale significantly larger than the distance.   In more formal terms, our partition is a special example of the general ($U(2)$ family of) point singularities allowed quantum mechanically on a line, whose realizations by scaled families of regular potentials has been studied extensively \cite{AGHH}.

Once the partition wall is realized and placed in the well separating the same number $N$ of particles on its two sides, under the finite temperature $T$ we expect that the wall is pushed from the two sides by the particles in motion.  Now, the point is that the different boundary conditions imposed at the partition cause different energy levels and, accordingly, different statistical distributions of particles in the levels between the two half wells, yielding 
a net force $\Delta F$ on the wall.  In \cite{FTjpa} we investigated the behaviour of the force $\Delta F$ as a function of $T$, and examined how it scales with particle number $N$ for the two kinds of particle statistics, the Bose-Einstein statistics and the Fermi-Dirac statistics. 
There we have found that the force $\Delta F$ has a finite limit at zero temperature $T = 0$ which scales as $N$ for bosons and as $N^2$ for fermions, and that it has a minimum before it diverges as $\sqrt{T}$ for the high temperature limit $T \to \infty$.

\begin{figure}[t]
\vskip 3.4ex
\centering
\resizebox{.4\textwidth}{!}{\includegraphics{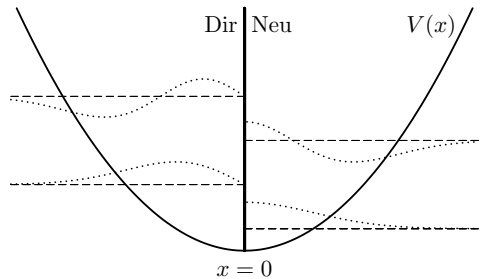}}
\caption{
The eigenfunctions and eigenvalues under the harmonic potential  $V(x) = \frac{m \omega^2}{2} x^2$ 
with a partition at the centre.  If the partition admits the Dirichlet ($ \psi
= 0 $) and the Neumann ($ \psi' = 0 $) boundary conditions on the left and on
the right, respectively, the eigenstates (the lowest two are shown in both half
lines) possess different energy levels.  When the same number $N$ of
particles are introduced in each of the half lines, these level
differences give rise to a net force on the partition.
}\label{fig:1}
\end{figure}

\begin{figure}[tb]
\vskip 3.4ex
\centering
\resizebox{.35\textwidth}{!}{\includegraphics{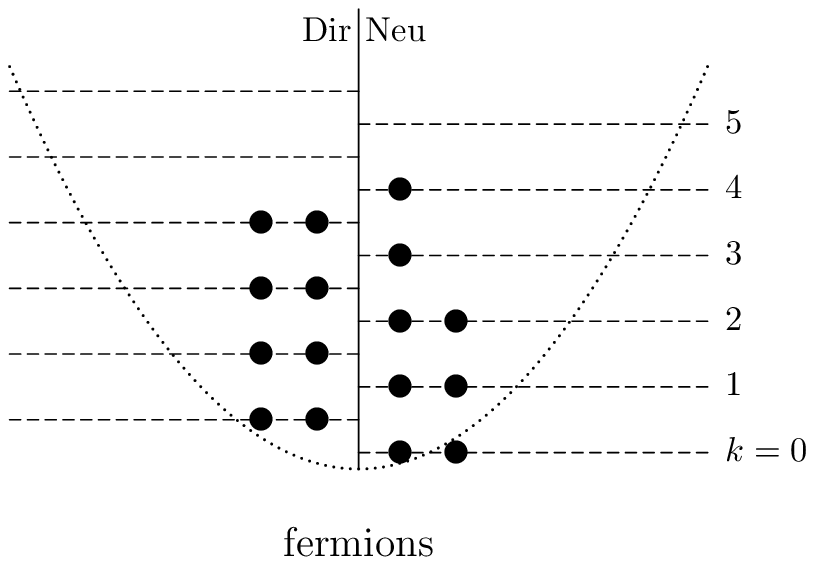}}
\hskip .1\textwidth
\resizebox{.35\textwidth}{!}{\includegraphics{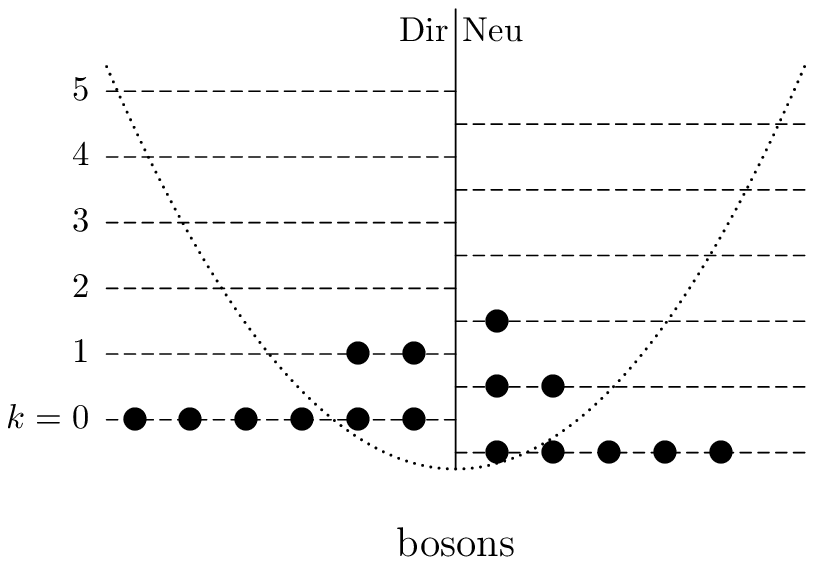}}
\caption{
Illustration of particle distributions over the levels in the two
half lines at a low temperature. The force related to each level differs
on the two sides of the partition and induces a non-vanishing net force
on the partition, which is dependent on temperature as well as on the
particle statistics.
}\label{fig:2}
\end{figure}

This raised a natural question if these results are specific to the potential used in the analysis, and if so how.  
In the present paper, we attempt to answer this by considering the same partition wall placed in a different potential, namely, the harmonic oscillator potential (see Figure~\ref{fig:1}).  One of the reasons for the harmonic potential is that, unlike the infinite well potential, it stretches infinitely for higher energies and shares a feature with potentials which are often used to describe actual physical systems.   Another reason is that,  on account of its technical simplicity which we also exploit here, the harmonic potential itself is widely used in various physical contexts including confinement of particles in a narrow region.    
Again, as illustrated in Figure~\ref{fig:2}, we consider both bosonic and fermionic cases for particle statistics, and the number $N$ of particles is
regarded as large but not macroscopically large (to ensure its possible relevance to nano devises).
We derive analytic approximate formulae for the force in the low, medium and high-temperature regimes separately, which are found to be reasonably good to reproduce the numerical results obtained for  $N = 100$.    
Note that, in actual realizations, our one dimensional system can be regarded as a model of an axis in three dimensions perpendicular to the surface of the partition which attracts the particles by the harmonic potential.    

Our analysis then shows that, in contrast to the infinite well case,  the force $\Delta F$ on the wall decreases as $1/\sqrt{T}$ as the temperature increases in the high temperature regime, and eventually vanishes in the limit $T \to \infty$.   This is the case for both bosons and fermions, and the force is of the order of $N$.  For the low temperature regime, the zero temperature limit of the force is of the order of $N$ for bosons as in the potential well case, but for fermions it is of the order of $\sqrt{N}$ in contrast to $N^2$ obtained in the potential well.   We also find that, unlike the infinite well case, no minimal point of the net force is found in the medium-temperature regime irrespective of the statistics of the particles.   Implications of these results, combined with the previous ones, will be discussed in the text.

This paper is organized as follows.  In section~\ref{s2} we define the model and provide our scheme of 
analytical approximation for the induced force at arbitrary finite temperatures.  
In section~\ref{s3}, we present our analysis in detail for the high temperature regime as well as the numerical results obtained.
Section~\ref{s4} is devoted to the analysis of the
low temperature regime, where we employ independent approaches for the  
fermionic and bosonic cases.
Section~\ref{s5} discusses the medium-temperature regime where we also seek  to interpolate the other two temperature regimes.   
Finally, we present our conclusion and discussions in section~\ref{s6}.

\section{Definition and basic properties of the system}
\label{s2}

In this section, we first define the system mentioned earlier and introduce some notations convenient for the description of the system.   With these, we present our basis of evaluating the induced force on the wall, and provide our scheme of analytical approximation used later.

\subsection{The eigenstates and distributions at finite temperatures}

To begin with, we recall the basic result of a harmonic oscillator system.  
{}For a particle moving on a line under the  
harmonic potential $V(x) = \frac{m \omega^2}{2} x^2$,  
the normalized energy
eigenfunctions $\phi_{l}(x)$ and the corresponding energy eigenvalues $\cE_l$
obeying the equations,
  \begin{equation}
  H\,  \phi_{l}(x) = \cE_l  \,  \phi_{l}(x),
  \qquad 
  H =  - \f{\h^2}{2m} \f{d^2}{dx^2} + \f{m \omega^2}{2} x^2,
    \label{eigeneq}
  \end{equation}
are given by
  \begin{equation}
  \!\!\!\!\!\!\!\!\! \!\!\!\!\!\!\!\!\!
  \phi_{l} (x) = \f{1}{\s{a}} \f{1}{\sqrt[4]{\pi}} \f{1}{\s{2^l l!}}
  \, \qH_l \! \pp1{ \f{x}{a} } \e{- \f{x^2}{2 a^2}}
  , \qquad
  \cE_l = \f{\cE}{2} \pp1{ l + \f{1}{2} }
  , \qquad l = 0, 1, 2, \ldots.
  \label{eigenfunc}
  \end{equation}
Here, $\qH_l(z)$ are the Hermite functions \cite{AS} and we have introduced 
the length scale  $a = \s{\f{\h}{m \omega}}$ and the
energy scale $\cE = 2 \h \omega = 2 \f{\h^2}{m a^2}$
(the unusual factor 2 is for our later convenience).
The Hermite functions are even functions for even $l = 2k$ and odd functions for odd $l = 2k + 1$, and fulfill
  \begin{equation}
  \qH_{2k} (0) = (-1)^k \f{(2k)!}{k!},
\qquad  
\qH_{2k + 1}' (0) = 2 (2k + 1) \qH_{2k} (0) .
\label{hzero}
  \end{equation}

Now let us divide the line into two by inserting an infinitely thin wall at $x = 0$ and consider its consequence in one of the half lines, say, the positive one.
If the boundary condition on the wave functions at the wall 
is the Dirichlet condition, then for the particle confined in the positive half line, we have only the
odd eigenfunctions, that is,
  \begin{equation}
  \varphi_k\qD = \s{2}\, \phi_{2k + 1} ,
  \qquad E_k\qD = \cE_{2k + 1}, \qquad k = 0, 1, \ldots,
  \label{diref}
  \end{equation}
where the factor $\s{2}$ is required for the normalization in the half line.
On the other hand, if it is the Neumann boundary condition, then we have only the even eigenfunctions,
  \begin{equation}
  \varphi_k\qN = \s{2}\, \phi_{2k} ,
  \qquad E_k\qN = \cE_{2k} , \qquad k = 0, 1, \ldots
    \label{newef}
  \end{equation}
One may wonder if there exist eigenfunctions other than these (\ref{diref}) and (\ref{newef}) in our system, since after all our partition wall breaks the parity symmetry under $x \to -x$ forcing us to work in two half lines separately,  and there seems no reason to consider only the standard eigenfunctions of the harmonic oscillator defined on the whole line with definite parities.   However, this cannot be the case, because if $\psi_+(x)$ is any eigenfunction defined on the positive half line obeying, say, the Dirichlet condition at $x = 0$, then one can extend it to the whole line by setting $\psi(x) = \psi_+(x)$ for $x \ge 0$ and $\psi(x) = -\psi_+(-x)$ for $x < 0$.   The resultant function $\psi(x)$ is regular at $x = 0$ and satisfies the equation (\ref{eigeneq}) on the whole line and hence provides an eigenfunction with odd parity of the harmonic oscillator, implying that it should be one of the standard ones $\phi_{2k + 1}$.   The case of the solutions with the Neumann condition can be argued similarly.

To proceed, consider a statistical system consisting of $N$ mutually noninteracting identical particles in the harmonic potential in the half line under finite temperature $T$.   For convenience, we introduce the dimensionless quantities,
  \begin{equation}
  t = k_{\rm B} T / \cE , \qquad b = 1/t ,
  \label{resctemp}
  \end{equation}
with $k_{\rm B}$ being the Boltzmann constant, and write the energy eigenvalues  as  
\begin{equation}
E_k = \cE e_k, \qquad \hbox{with} \quad  e_k = (k + \sigma) , \qquad k = 0, 1, \ldots,   
\end{equation}
where the constant $\sigma$ is given, for the Dirichlet case and the Neumann case, by 
  \begin{equation}
 \sigma\qD = \f{3}{4} , \qquad \sigma\qN = \f{1}{4},
  \end{equation}
respectively.
The statistical distributions of the particles then read
  \begin{equation} \label{hae}
  N_k = \f{1}{ \e{\alpha + b e_k} - \eta } =
  \f{1}{ \e{(\alpha + b \sigma) + b k} - \eta } , 
\qquad
 \eta = \left\{ 
  \begin{array}{ll}
  1 & \mbox{bosons} \\
  -1 & \mbox{fermions}
  \end{array}
  \right. 
\end{equation}

At a given temperature, the chemical constant $\alpha$ is determined uniquely from the constraint
$N = \sum_{k=0}^{\infty} N_k$.  This implies that the combination $\alpha + b \sigma$ is uniquely determined by the  total number constraint, from which we learn that the distributions are actually the same for both of the Dirichlet and the Neumann boundary conditions, that is, $ N_k\qD = N_k\qN$ for all $k$.  
This allows us to introduce
 \begin{equation} 
 \qa := \alpha\qD + b \sigma\qD = \alpha\qN + b \sigma\qN, \qquad
N_k := N_k\qD = N_k\qN
  \end{equation}
to obtain the simpler expression,
  \begin{equation}  \label{hal}
  N_k =  \f{1}{ \e{\qa + b k} - \eta }  \qquad
  k = 0, 1, \ldots,
  \end{equation}
commonly used for the two boundary conditions.
We also find from $1/N_k = \e{\qa + b k} - \eta$ and $b > 0$ that 
$1/N_0 < 1/N_1 < 1/N_2 < \ldots$, or
$N_0 > N_1 > N_2 > \ldots$.
Note that for bosons $\eta = 1$, the positivity of distributions $N_k > 0$ for all $k$ implies $\qa > 0$ at any temperature (where the case $k=0$ gives 
the strongest condition).  For fermions $\eta = -1$, no such restriction emerges and
$\qa$ can take any value in $(-\infty, \infty)$.

At this point, we mention that the sum over the levels $k$ admits an exact resummation valid
for $\qa > 0$,
  \begin{eqnarray} \label{hah}
  N & = & \sum_{k=0}^{\infty} \f{1}{ \e{\qa + b k} - \eta } =
  \sum_{k=0}^{\infty} \f{\e{-(\qa + b k)}}{1 -
  \eta \e{-(\qa + b k)} } = \eta \sum_{k=0}^{\infty}
  \f{\eta \e{-\qa} \e{-b k}}{ 1 - \eta \e{-\qa} \e{-b k} }
  \nonumber \\ 
  & = & \eta \sum_{k=0}^{\infty} \sum_{l=1}^{\infty}
  \pp1{ \eta \e{-\qa} \e{- b k} }^l =
  \eta \sum_{l=1}^{\infty} \pp1{ \eta \e{-\qa} }^l
  \sum_{k=0}^{\infty} \e{- b k l}
  =\eta \sum_{l=1}^{\infty}
  \f{\pp1{ \eta \e{-\qa} }^l}{1 - \e{-bl}} \ ,
  \end{eqnarray}
where we have used $\eta^{-1} = \eta$.
This resummation formula will be useful later.

\subsection{The force difference} \label{sec:hac}

Let us now suppose that the wall inserted at $x = 0$ in the harmonic
potential imposes the Dirichlet boundary condition on the left (negative)
side and the Neumann boundary condition on the right (positive) side.
Due to the difference in the energy levels developed in the two sides of
the wall, one expects that a net force, or statistical pressure, will
emerge on the wall as a purely quantum effect deriving from the boundary
conditions (see Figure~\ref{fig:2}). Our aim is to evaluate this induced net force as a function
of (rescaled) temperature $t$.

Before we proceed, we 
recall the fact that for the case of the infinite potential well \cite{FTjpa}, the force acting on the wall from each side of the half lines 
proves to be essentially the same as the one giving the average energy, 
$\bar E = \sum_{k=0}^{\infty} N_k {E_k}$.  From this, the net force is obtained by
the difference $\Delta \bar E = \bar E\qD - \bar E\qN$.
For the harmonic oscillator, the difference in the average energy reads  
\begin{eqnarray}
\Delta \bar E 
  = \sum_{k=0}^{\infty} N_k \pp1{ E_k\qD - E_k\qN }
  = \f{\cE}{2}\sum_{k=0}^{\infty} N_k \pp1{ \sigma\qD - \sigma\qN } = \f{N\cE}{2} ,
  \end{eqnarray}
which is temperature-independent.

However, for the harmonic oscillator the average energy is no longer the same as the force.  To see this, 
let us consider the contribution $F$ for the total force coming from one single level $E$.  Under a shift $\delta x$ of the wall from the origin, the energy level will also change by $\delta E$, and from this the force is found by 
  \begin{equation}
   \label{forcedef}
F =  - \lim_{\delta x \to 0} \f{ \delta E }{ \delta x }.
  \end{equation}
The total net force can then be obtained by gathering the force difference $\Delta F_k$ for all $k$, which is the difference of the forces between the two sides of the wall arising from the two corresponding energy levels specified by the same integer $k$.  
Unfortunately, unlike the infinite well case, we do not have analytical solutions for the harmonic oscillator when the partition is displaced from the centre, and we are compelled to resort to some approximation scheme to evaluate the force (\ref{forcedef}).

At this point we recall that, mathematically speaking, the Hamiltonian operator $H$ of our system has 
the infinity $x = \pm \infty$ as a limit-point
singularity whereas the position of the wall is a regular endpoint, meaning that  for any real eigenvalue $E$, there exists only one normalizable
eigenfunction up to a phase
factor.   In more concrete terms, given an arbitrary $E$ we have two independent  solutions for the differential equation (\ref{eigeneq}) but requirement of normalizability 
determines a particular linear combination of the two as a possible candidate for an eigenfunction.   It qualifies as a true eigenfunction when the boundary condition
at the partition is further met, which is attained by tuning $E$ to be one of the particular set of real numbers which form the energy spectrum of the system.   
This heuristic picture of approaching eigenfunctions and eigenvalues suggests that,
if the shift $\delta x$ of the wall is sufficiently small,  for a fixed (Dirichlet or Neumann) boundary condition
the difference in the 
eigenfunctions should be small
in the $L^2$ sense that  
their scalar product tends to 1 as $\delta x \to 0$,
with the perturbed eigenvalue $\tilde E$ also being close to the unperturbed one
$E$.  
Similarly, we expect that for a small variation $\delta x$ 
the difference in the boundary values of the two wave functions or their derivatives --
the former is nonvanishing for the Neumann case while the latter is nonvanishing for the Dirichlet  case -- 
remains small and in the same order $\Ordo{\delta x}$ at most.   

Now, we consider the identity
valid for any two real and normalized eigenfunctions,
 \begin{eqnarray} 
  \pp1{ E- \tilde E } \pp1{ \varphi_{\tilde E}, \varphi_{ E} }\qplus 
  &=&
  \pp1{ \varphi_{\tilde E}, H \varphi_{ E} }\qplus -
  \pp1{ H \varphi_{\tilde E}, \varphi_{ E} }\qplus
    \nonumber \\ 
  &=&
  \f{\h^2}{2m}
  \pp2{ \varphi_{\tilde E} (0) \, \varphi_{ E}' (0) -
  \varphi_{\tilde E}' (0) \, \varphi_{ E} (0) } ,
 \label{ibp}
  \end{eqnarray} 
where $\pp1{ \, \cdot \, , \, \cdot \, }\qplus$ denotes the scalar
product on the positive half line, and the prime indicates the derivative with respect to $x$, {\it e.g.}, $\varphi' = \f{d}{dx}\varphi$.
Specifically, for the Dirichlet case,  we choose in (\ref{ibp}) an unperturbed Dirichlet eigenvalue $E\qD_k$ and its eigenfunction $\varphi\qD_k$
for $E$ and $\varphi_{E}$, and the perturbed eigenvalue and eigenfunctions caused by the shift in the wall for 
$E$ and $\varphi_{E}$, respectively.
Based on our observations on the perturbed quantities, we find that the formula (\ref{ibp}) in the leading
order of $\delta x$ or of $\delta E = E - E\qD_k$ yields
  \begin{equation}
  \delta E = E - E\qD_k \approx - \f{\h^2}{2m}
  { \varphi_k\qD }{}' (0) \, \varphi_{E} (0),
  \end{equation}
or
  \begin{equation}
  \varphi_{E} (0) \approx
  - \f{ \delta E } { \f{\h^2}{2m} { \varphi_k\qD }{}' (0) } .
  \end{equation}
In parallel, our assumption ensures that 
  \begin{equation}
  { \varphi_{E} }' (0) \approx { \varphi_k\qD }{}' (0) ,
  \end{equation}
and that the Dirichlet condition is satisfied at the shifted wall $x = \delta x$,
  \begin{equation}
  0 = \varphi_{E} (\delta x) \approx
  \varphi_{E} (0) + { \varphi_{E} }' (0) \cdot \delta x ,
  \end{equation}
from which we have
  \begin{equation}
  \pp1{ \f{ \delta E } { \delta x } }_k\qD \approx
  \f{\h^2}{2m} \pp2{ { \varphi_k\qD }{}' (0) }^2 .
    \label{denergy}
  \end{equation}

Analogously, for the Neumann case, we have
  \begin{equation}
  \delta E = E - E\qN_k \approx \f{\h^2}{2m}
  \varphi_k\qN (0) \, \varphi_{E}{}' (0) ,
  \end{equation}
or
  \begin{equation}
  \varphi_{E}{}' (0) \approx
  \f{ \delta E } { \f{\h^2}{2m} \varphi_k\qN (0) } .
  \end{equation}
Since $\varphi_{E}$ is an eigenfunction, we find
  \begin{equation}
  {\varphi_{E}}'' (0) = - \f{2m E}{\h^2} \varphi_{E}(0)
  \approx - \f{2m E\qN_k}{\h^2} \varphi_k\qN (0) .
  \end{equation}
The Neumann condition is satisfied at $x = \delta x$ if
  \begin{equation}
  0 = {\varphi_{E}}' (\delta x) \approx
  \varphi_{E}' (0) + { \varphi_{E} }'' (0) \cdot \delta x ,
  \end{equation}
from which we obtain
  \begin{equation}
  \pp1{ \f{ \delta E } { \delta x } }_k\qN \approx
  E\qN_k \pp2{ { \varphi_k\qN } (0) }^2 .
  \label{nenergy}
  \end{equation}

Combining (\ref{denergy}) and (\ref{nenergy}) together with (\ref{eigenfunc}) and (\ref{hzero}), 
one can evaluate the contribution for the force difference coming from the $k$-th level as
  \begin{eqnarray}
  && \!\!\! \!\!\! \!\!\! \!\!\! \!\!\! \!\!\! \!\!\! \!\! \Delta F_k = \f{\h^2}{2m} \pp2{ { \varphi_k\qD }{}' (0) }^2
  - E\qN_k \pp2{ { \varphi_k\qN } (0) }^2
  \nonumber \\
  &&  \!\!\! \!\!\! \!\!\! \!\!\!  =  \f{\h^2}{2 m a^2} \f{2}{\s{\pi}} \f{1}{2^{2k + 1} (2k + 1)!}
  \pp2{ \qH_{2k + 1}' (0) }^2
- \cE \pp1{ k + \f{1}{4} } \f{2}{\s{\pi} a}
  \f{1}{2^{2k} (2k)!} \pp2{ \qH_{2k} (0) }^2
  \nonumber \\
  &&  \!\!\! \!\!\! \!\!\! \!\!\!  =
  \f{\cE}{ 2 \s{\pi} a } \f{ (2k)! }{ 2^{2k} (k!)^2 } . \nonumber
  \end{eqnarray}
The final expression shows that $\Delta F_k$ has a nontrivial $k$-dependence in contrast to the
infinite well case \cite{FTjpa} where it is simply proportional to $k$.

In what follows, for brevity we shall use the dimensionless
force difference, 
  \begin{equation}
  \Delta f_k :=  { 2 \s{\pi} a } {{\Delta F_k}\over{\cE}} = \f{ (2k)! }{ 2^{2k} (k!)^2 },
   \end{equation}
which has the first few values,  
\begin{equation} \label{hab}
  \Delta f_0 = 1, \qquad \Delta f_1 = \f{1}{2}, \qquad
  \Delta f_2 = \f{3}{8}, \qquad \Delta f_3 = \f{5}{16}.
  \end{equation}
Note that, in general, $\Delta f_k > \Delta f_{k+1}$
(since $ \Delta f_{k+1} / \Delta f_k = (2k+1)/(2k+2) < 1 $)
and, consequently, $\Delta f_k \le 1$.
To see how it behaves for  large $k$, we may use the Stirling formula \cite{ST}
  \begin{equation}
  n! = \pp1{ \f{n}{\ee} }^n \s{2 \pi n}
  \pp2{ 1 + \f{1}{12 n} + \Ordo{\f{1}{n^2}} }
  \end{equation}
to obtain
  \begin{equation} \label{hbt}
  \f{ (2k)! }{ 2^{2k} (k!)^2 } = \f{1}{\s{\pi k}}
  \pp2{ 1 - \f{1}{8k} + \Ordo{\f{1}{k^2}} } ,
  \end{equation}
which provides a sufficiently good approximation already at $k = 1$,
and improves quickly for larger $k$.
One can also observe that the \lq rearranged\rq\ approximation
 \begin{equation} \label{hbs}
  \Delta f_k \approx \f{1}{\s{ \pi \pp1{ k + \F{1}{4} } }}
 \end{equation}
is similarly good as (\ref{hbt}),
and can be used for $k=0$ as well (see Table~1).

\vskip 4ex
\begin{center}
  \begin{tabular}%
[t] %
{|r|rlrrcrrcrrcr|}  \hline
       \multicolumn{1}{|r|}{$\rule[-3.5ex]{0em}{7.5ex} k$}
    &  \multicolumn{3}{c}{$ \Delta f_k$}
    &  \multicolumn{3}{c}{$ \f{1}{\s{\pi k}}$}
    &  \multicolumn{3}{c}{$ \f{1}{\s{\pi k}}  \pp2{ 1 - \f{1}{8k} }$}
    &  \multicolumn{3}{c|}{$ \f{1}{\s{ \pi \pp1{ k + \F{1}{4} } }}$}
    \\  
    \hline
    0 & & 1
    &&& \multicolumn{1}{c}{}  &&&  \multicolumn{1}{c}{}
    &&&  1.12838  &  \\  \hline
    1 & & 0.5    &&& 0.56419 &&& 0.49367 &&& 0.50463 &  \\  \hline
    2 & & 0.375  &&& 0.39894 &&& 0.37401 &&& 0.37613 &  \\  \hline
    3 & & 0.3125 &&& 0.32574 &&& 0.31216 &&& 0.31296 &  \\  \hline
  \end{tabular}%
\end{center}
\vskip 1ex
\centerline{Table 1: Exact and approximate values for $\Delta f_k$, for
$k = 0, 1, 2, 3$.}
\vskip 6ex

The total force difference (the net
force) is then given by
  \begin{equation} \label{haf}
  \Delta f = \sum_{k=0}^{\infty} N_k \, \Delta f_k =
  \sum_{k=0}^{\infty} \f{ (2k)! }{ 2^{2k} (k!)^2 }
  \f{1}{ \e{\qa + b k} - \eta } .
  \end{equation}
Now, using the resummation formula analogous to (\ref{hah}) together with the fact that a sum of the form
 \begin{equation}
  \sum_{k=0}^{\infty} \f{ (2k)! }{ 2^{2k} (k!)^2 } \, q^k =
  \sum_{k=0}^{\infty} \pp1{ 2k \atop k } \pp1{\f{q}{4}}^k
 \end{equation}
is actually the Taylor expansion of $ \; \F{1}{\s{1 - q}} \; $
for $ { |q| < 1 } $, we finally arrive at the convenient analytical expression of the net force:
 \begin{equation} \label{hag}
  \Delta f = \eta \sum_{l=1}^{\infty} \pp1{ \eta \e{-\qa} }^l
  \sum_{k=0}^{\infty} \f{ (2k)! }{ 2^{2k} (k!)^2 } \e{- b k l}
  =\eta \sum_{l=1}^{\infty}
  \f{\pp1{ \eta \e{-\qa} }^l}{\s{1 - \e{-bl}}} .
 \end{equation}

We mention that the inequality $\Delta f_k
\le 1$ noted earlier implies
 \begin{equation} 
  \Delta f = \sum_{k=0}^{\infty} N_k \, \Delta f_k \le
  \sum_{k=0}^{\infty} N_k = N.
   \label{ham}
 \end{equation}
Clearly, this inequality (\ref{ham}) is expected to be close to the equality for low temperatures but
it will become loose as temperature increases where the energy levels of higher $k$ are
excited with increasing probability.   This way we can expect
that the net force decreases as temperature increases.
One can also obtain an improved inequality utilizing the closer details $\Delta f_0 = 1$ and
$\Delta f_k \le 1/2$ for $k>0$, that is,
 \begin{eqnarray}  \label{han}
  \!\!\! \!\!\! \!\!\! \!\!\!     \!\!\! \!\!\! \!\!\! \!\!\! 
  \Delta f = \sum_{k=0}^{\infty} N_k \, \Delta f_k 
  = N - \sum_{k=1}^{\infty} N_k (1 - \Delta f_k)  
\le N - \sum_{k=1}^{\infty} N_k \cdot 1/2 
  =
 (N + N_0)/2.
 \end{eqnarray}

We have furnished analytical approximations for the net force which form our basis for studying the temperature dependence of the force on the partition wall.   We stress that, unlike the case of the potential well, 
we further need to improve the approximations to obtain results comparable to numerical computations.   
This is required by the nontrivial level dependence (\ref{haf}) of the force in the harmonic case, and below we shall establish independent approaches to deal with infinite sums which are appropriate for three different temperature (high, low and medium) regimes.

\section{The high-temperature regime}
\label{s3}

We now analyze the behavior of the net force when the temperature is sufficiently high 
$t \gg 1$ or $b \ll 1$ (recall that $t = 1/b$ is the rescaled dimensionless temperature in (\ref{resctemp})), where we are allowed to take $\qa > 0$.   
One can then approximate the formula (\ref{hah}) as
 \begin{equation} \label{hai}
  N = \eta \sum_{l=1}^{\infty} \f{\pp1{ \eta \e{-\qa} }^l}{1 - \e{-bl}}
  \approx \eta \sum_{l=1}^{\infty} \f{\pp1{ \eta \e{-\qa} }^l}{bl} \, .
 \end{equation}
This approximation is certainly good for the terms $l \ll t$, but if we
assume the condition $\e{-\qa} \ll 1$ (which is stronger than $\qa > 0$) for which 
higher-$l$ terms in the sum (\ref{hai})
are suppressed, then we can
rewrite (\ref{hai}) and perform the summation in closed
form as
 \begin{equation} \label{haj}
  \eta \f{N}{t} \approx \sum_{l=1}^{\infty}
  \f{\pp1{ \eta \e{-\qa} }^l}{l} = \ln \f{1}{ 1 - \eta \e{-\qa} } ,
 \end{equation}
which implies
 \begin{equation} \label{hak}
  \e{-\qa} \approx \eta \pp1{ 1 - \e{-\eta \f{N}{t}} } = \f{N}{t} -
  \f{\eta}{2} \pp1{ \f{N}{t} }^2 + \f{1}{6} \pp1{ \f{N}{t} }^3 +
  \Ordo{ \pp1{ \f{N}{t} }^4 } .
 \end{equation}
We can see from this outcome that for sufficiently high temperatures $t \gg N$ our approximation is indeed
consistent with $\e{-\qa} \ll 1$, suggesting that a valid approximation for $\qa(t)$ can be obtained at least for 
$t$ with $t \gg N$.  Based on this observation, we shall define the high-temperature regime of the system by the condition
$t \gg N$.  Notably, here $\qa$ depends on $t$ and $N$ only
through the combination $t/N$.

\begin{figure}[t]
\vskip 4ex
\centering
\resizebox{.34\textwidth}{!}{\includegraphics
{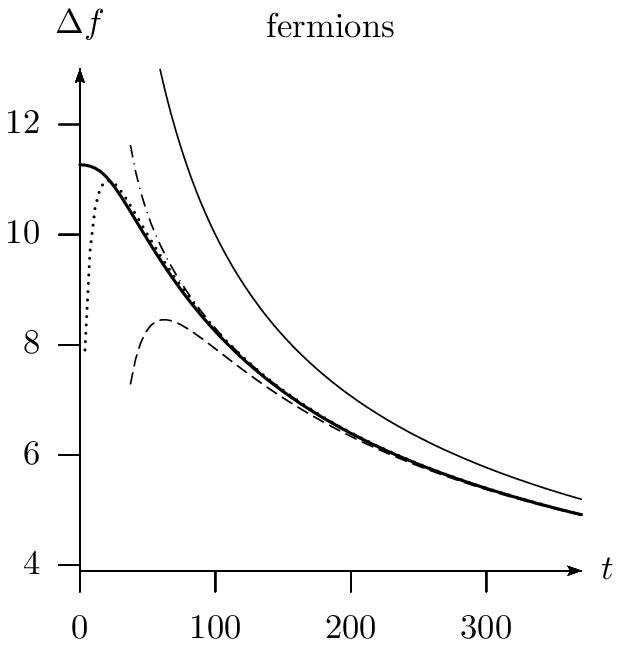}}  \hspace{10mm}
\resizebox{.35\textwidth}{!}{\includegraphics
{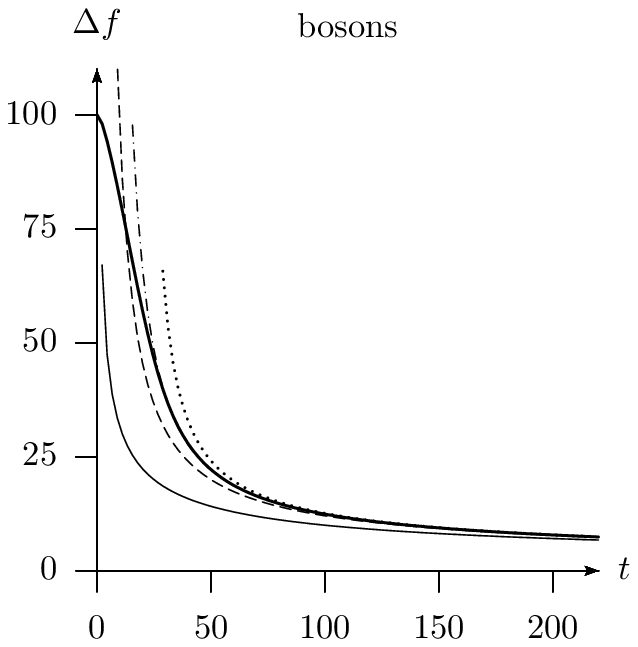}} 
\caption{
High temperature approximations for $\Delta f$, for fermions (left) and
bosons (right), $N=100$. Thick solid line: the numerical result. Thin
solid line: the expansion (\ref{hat}) truncated to one term. Dashed
line: truncation to two terms. Dash-dotted line: truncation to three
terms. Dotted line: (\ref{hax}). 
}\label{hbg}
\end{figure}

Similarly, we can also approximate the net force (\ref{hag}) as
\begin{equation} \label{hao}
\!\!\! \!\!\! \!\!\! \!\!\! \!\!\! 
  \Delta f = \eta \sum_{l=1}^{\infty} \f{\pp1{ \eta \e{-\qa} }^l}
  {\s{1 - \e{-bl}}} \approx \eta \sum_{l=1}^{\infty} \f{\pp1{ \eta
  \e{-\qa} }^l}{\s{bl}} = \eta \s{N} \pp1{\f{t}{N}}^{\f{1}{2}}
  \sum_{l=1}^{\infty} \f{\pp1{ \eta \e{-\qa} }^l}{\s{l}} \, .
 \end{equation}
Inserting the expansion (\ref{hak}), we find
\begin{equation} \label{hat}
\!\!\! \!\!\! \!\!\! \!\!\! \!\!\! \Delta f 
  \approx \s{N} \pp1{ \f{N}{t}}^{\f{1}{2}}
  \left[ 1 + \eta\, c_1 \f{N}{t} + c_2
  \pp1{\f{N}{t}}^2
+ \eta\, c_3
  \pp1{\f{N}{t}}^3 + \Ordo{ \pp1{ \f{N}{t} }^4 } \right] ,
 \end{equation}
with the coefficients 
\begin{equation}
\!\!\! \!\!\! \!\!\! \!\!\! 
c_1 =  \f{\s{2} - 1}{2}, \qquad c_2 =  \f{1 - 3\s{2} + 2\s{3}}{6}, \qquad  c_3 = \f{11 + 7\s{2} - 12\s{3}}{24}.
\end{equation}
Their numerical values are estimated as
\begin{equation}
  c_1 \approx  0.207, \qquad c_2   \approx 0.0369 ,  \qquad c_3   \approx 0.00479. 
\end{equation}

\begin{figure}[t]
\vskip 4ex
\centering
\resizebox{.37\textwidth}{!}{\includegraphics
{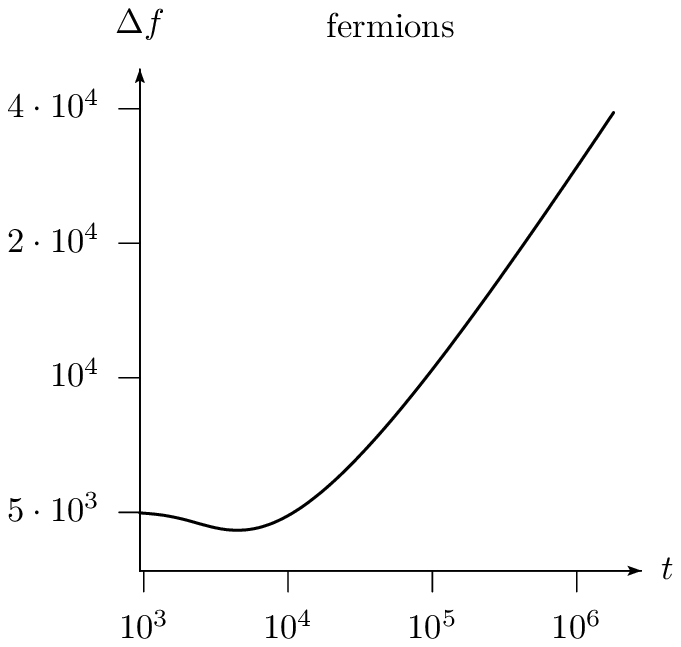}}  \hspace{10mm}
\resizebox{.35\textwidth}{!}{\includegraphics
{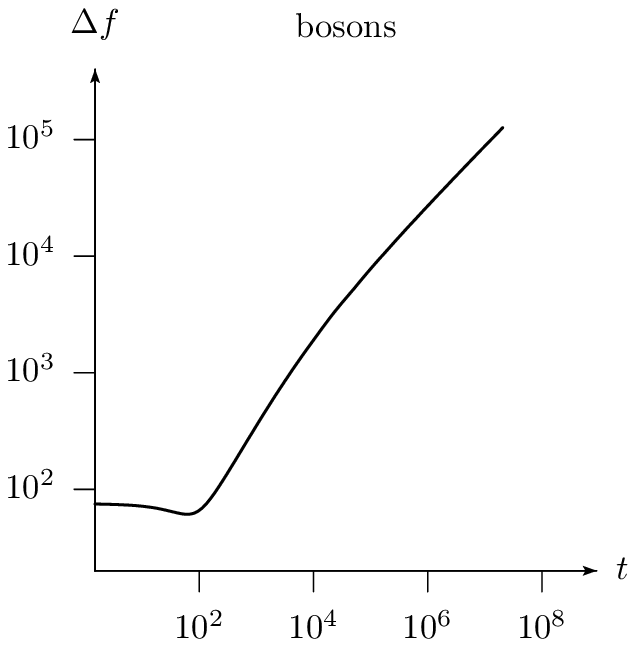}} 
\caption{
Numerical results of the net force $\Delta f$ in the infinite well case; for fermions (left) and
bosons (right), at $N=100$.   The net force diverges as $t \to \infty$ according to $\sqrt{t}$ for both fermions and bosons \cite{FTjpa}.
}\label{addedfigure}
\end{figure}

The analytical results, together with those improved below, are depicted in Figure~\ref{hbg} along with the numerical ones.  There we find that, for both of the fermionic and the bosonic cases,  the net force is a monotonically decreasing function in the high-temperature regime and vanishes in the limit $t \to \infty$.  In fact, the expression (\ref{hat}) shows that 
it decreases according to $1/\sqrt{t}$ asymptotically as $t \to \infty$ with the common order $N$.  This outcome is in sharp
contrast to the infinite well case \cite{FTjpa} where the force $\Delta f(t)$ diverges according to $\sqrt{t}$ for both fermions and bosons (see Figure~\ref{addedfigure}).

It is worth noting that this high-temperature expansion can be made applicable even for medium
temperatures if we modify it slightly by adopting the Pad\'e approximant form \cite{BG}  to regularize
its diverging behavior for $t \to 0$.  Explicitly, we may take
 \begin{equation} \label{hax}
  \Delta f \approx \s{N} \f{ \pp1{\f{N}{t}}^{\f{1}{2}} }
  { 1 - \eta \f{\s{2} - 1}{2} \f{N}{t} } ,
  \hskip 1.9em  \hbox{or}  \hskip 1.9em
  \Delta f \approx \s{N} \f{ \pp1{\f{t}{N}}^{\f{1}{2}} }
  { \f{t}{N} - \eta \f{\s{2} - 1}{2} }
 \end{equation}
both of which admit the expansion (\ref{hat}) but with slightly modified coefficients,
\begin{equation}
  c_1 \approx  0.207, \qquad c_2   \approx 0.0429 ,  \qquad c_3   \approx 0.00888. 
\end{equation}
We note that the regularized formula (\ref{hax}) reproduces the analytical approximation 
(\ref{hat}) quite well up to the third term (the fourth also not being very different).
However, for bosons the attempt to regularize at $t=0$ fails because
(\ref{hax}) diverges at a certain positive medium temperature value, {\it i.e.},
$\f{t}{N} = \f{\s{2} - 1}{2}$, and this also ruins improvement in
precision.  For fermions, in contrast, the approximation is indeed valid up to fairly low temperatures.
It is also important to observe from the result (\ref{hat}) that, similarly to
$\qa$, the net force $\Delta f/\s{N}$ also depends on $t$ and $N$ only  
through the combination $t/N$ in the high-temperature regime.

In passing, we provide a technical remark that, if one wishes to improve the approximations
(\ref{hai}) and (\ref{hao}) to curb the error caused by the replacement
$1 - \e{-bl} = bl + \Ordo{ (bl)^2 } \approx bl$ used above, one may instead use
$ \e{\f{bl}{2}} - \e{-\f{bl}{2}} = bl + \Ordo{ (bl)^3 } \approx bl $ (which is more precise by one order) to obtain 
 \begin{equation} \label{har}
 \!\!\! \!\!\! \!\!\! \!\!\! 
  \sum_{l=1}^{\infty} \f{\pp1{ \eta \e{-\qa} }^l}{1 - \e{-bl}} 
  \approx
  \sum_{l=1}^{\infty} \f{\pp1{ \eta \e{-\qa+\f{b}{2}} }^l}{bl},
  \qquad
  \sum_{l=1}^{\infty} \f{\pp1{ \eta \e{-\qa} }^l}{\s{1 - \e{-bl}}}
  \approx
  \sum_{l=1}^{\infty} \f{\pp1{ \eta \e{-\qa+\f{b}{4}} }^l}{\s{bl}} \, ,
 \end{equation}
respectively.  Although the acquired improvements in $\qa$ and in $\Delta f$ are 
of the order of $1/t$ and hence insignificant for $N \to \infty$, they
may become significant for $N \sim 100$.

\section{The low-temperature regime}
\label{s4}

Next we turn our attention to the net force when the temperature is sufficiently low.    In this regime we need to develop our
approximation depending the statistics of the particles, and below we present our arguments for fermions and bosons, separately.

\subsection{Fermions at low temperature}

Note first that, for the fermionic case $\eta = -1$,  the particle distribution at $t=0$ becomes
 \begin{equation} \label{hba}
  N_0 = N_1 = \cdots = N_{N-1} = 1,  \qquad  N_N = N_{N+1} = \cdots = 0,
 \end{equation}
for which $\qa$ is, so to say, $-\infty$. 
The net force is then
 \begin{equation} \label{hbb}
 \!\!\! \!\!\! \!\!\! \!\!\! \!\!\! \!\!\! \!\!\! \!\!\! 
  \Delta f (0) = \sum_{k=0}^{N-1} \Delta f_k = \sum_{k=0}^{N-1}
  \f{(2k)!}{2^{2k} (k!)^2} = \f{2N (2N)!}{2^{2N} N!^2} =
  2 \s{\f{N}{\pi}} \pp2{1 - \f{1}{8N} + \Ordo{ {\f{1}{N^2}} }},
 \end{equation}
where the
exact result is, at the last step,
approximated for large $N$ using the Stirling formula.  Observe that the limiting value $\Delta f (0)$ is 
of the order $\sqrt{N}$ for large $N$.

For a slightly higher temperature $t > 0$, we may assume that 
the distribution differs from (\ref{hba}) only
at $N_{N-1}$ and $N_N$. 
Then $\sum_{k=0}^\infty N_k = N$ implies
$N_{N-1} + N_N = 1$, which can be exploited as
 \begin{equation}
  \f{1}{\e{ \qa + bN } + 1} = N_N = 1 - N_{N-1} = 1 -
  \f{1}{\e{ \qa + b(N-1)} + 1} =
  \f{\e{ \qa + b(N-1) }}{\e{ \qa + b(N-1)} + 1} 
 \end{equation} 
which is equivalent to
 \begin{equation} \label{hbc}
  \qa = -b \pp1{ N - \f{1}{2} } , 
 \end{equation}
in this approximation.

For the net force, we can use the same approximate distribution, but we may further suppose that
(\ref{hbc}) is a good approximation even when more than two
levels --- let their number be denoted by $2J$, still assuming $2J \ll
2N$ --- get nontrivially occupied.
Then the net force $\Delta f$ will differ from the zero
temperature value $\Delta f(0)$ as
\begin{eqnarray}
&& \!\!\! \!\!\! \!\!\! \!\!\! \!\!\!  
\Delta f - \Delta f(0) 
  = \sum_{k=0}^\infty \Delta f_k N_k -
  \sum_{k=0}^{N-1} \Delta f_k = - \sum_{k=0}^{N-1} \Delta f_k
  \pp1{1 - N_k} + \sum_{k=N}^\infty \Delta f_k N_k
  \nonumber \\ 
&& \!\!\! \!\!\! \!\!\! \!\!\! \!\!\!    
\approx - \sum_{k=N-J}^{N-1} \Delta f_k \pp1{1 - N_k} +
  \sum_{k=N}^{N+J-1} \Delta f_k N_k
    \nonumber \\ 
&& \!\!\! \!\!\! \!\!\! \!\!\! \!\!\!  
  = - \sum_{j=1}^J \Delta f_{N-j} \pp1{1 - \f{1}{ \e{b
  \pp1{\f{1}{2} - j}} +1 }} + \sum_{j=1}^J \Delta f_{N-1+j}
  \f{1}{ \e{b \pp1{j - \f{1}{2}}} +1 }.
\end{eqnarray}
This can be further approximated by using (\ref{hbs}) as 
\begin{eqnarray}
  \Delta f - \Delta f(0) 
  = \sum_{j=1}^J \f{ \Delta f_{N-1+j} - \Delta f_{N-j}}
  { \e{b \pp1{j - \f{1}{2}}} +1 } 
  \approx \sum_{j=1}^J
  \f{ \f{1}{ 2\s{\pi} N^{\f{3}{2}} } \pp1{1 - 2j} 
  }{ \e{b \pp1{j - \f{1}{2}}} +1 },
\end{eqnarray}
from which we obtain
 \begin{equation} \label{hbd}
  \Delta f \approx \Delta f(0) - \f{1}{ 2\s{\pi} N^{\f{3}{2}} } \pp2{
  \f{1}{ \e{\f{b}{2}} +1 } + \f{3}{ \e{\f{3b}{2}} +1 } + \cdots },
 \end{equation}
on account of $j \le J \ll N$. 
As an expansion in terms of $\e{-\f{b}{2}}$,
(\ref{hbd}) turns into
 \begin{equation} \label{hbe}
  \Delta f \approx \Delta f(0) - \f{1}{ 2\s{\pi} N^{\f{3}{2}} } \pp2{
  \e{-\f{b}{2}} - \e{-b} + 4 \e{-\f{3b}{2}} - \e{-2b} +
  6 \e{-\f{5b}{2}} + \cdots }.
 \end{equation}
This result shows that, relative to the zero temperature value whose leading order is $\sqrt{N}$ (see (\ref{hbb})), the temperature dependence of  the net force $\Delta f$ starts at the order of $1/N^2$.
As a result, a noticeable deviation from
$\Delta f(0)$ can be expected only for $t \gg 1$.

\begin{figure}[bt]
\vskip 4ex
\centering
\resizebox{.38\textwidth}{!}{\includegraphics{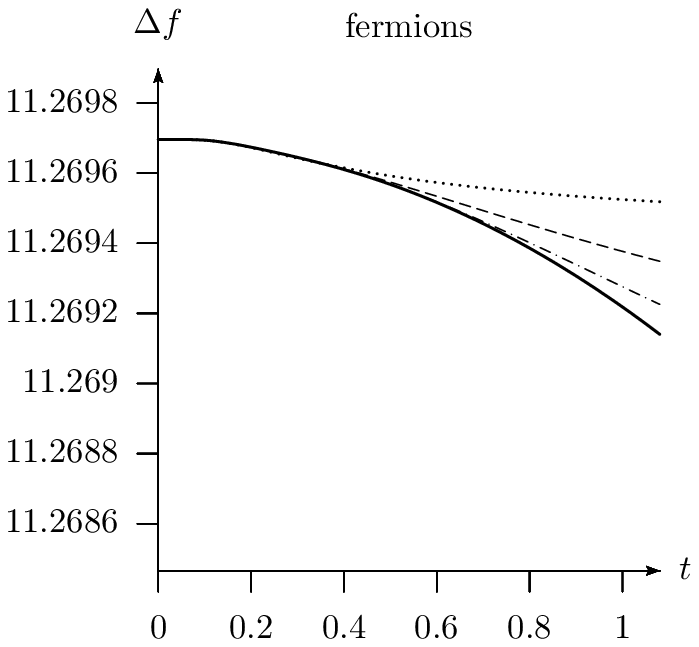}}  \hspace{10mm}
\resizebox{.365\textwidth}{!}{\includegraphics{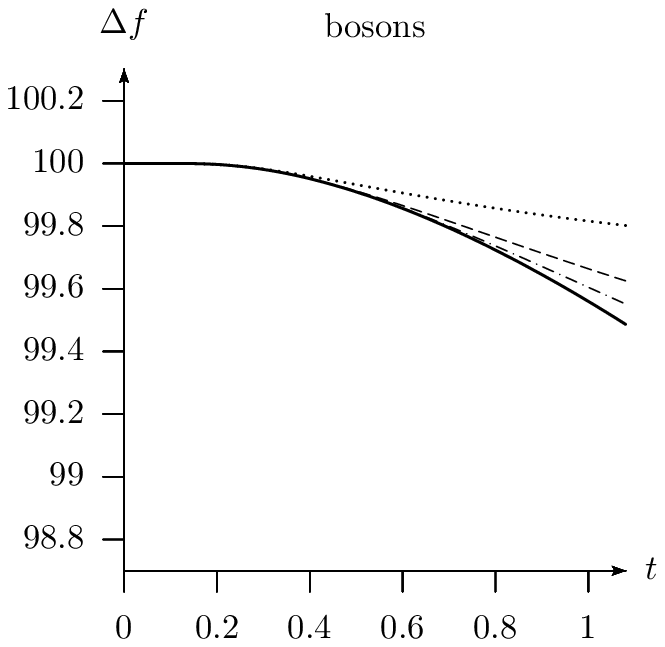}}
\caption{
Low temperature approximations for $\Delta f$ for fermions (left) and
bosons (right), $N=100$. Thick solid line: the numerical result. Dotted
line: the fermionic expansion (\ref{hbe}) up to the term
$\e{-\f{b}{2}}$, respectively the bosonic expansion (\ref{hbl}) up
to the term $\e{-b}$. Dashed line: terms up to $\e{-\f{3b}{2}}$, resp.\
$\e{-2b}$. Dash-dotted line: terms up to $\e{-\f{5b}{2}}$, resp.\
$\e{-3b}$.
}\label{hbf}
\end{figure}

\subsection{Bosons at low temperature}

{} For the bosons $\eta = 1$, we know that 
at $t = 0$, the distribution is given by $N_0 = N$ and $N_k = 0$ for $k>0$, and
hence $ N_0 = 1/{\e{\qa} - 1} = N $.  This implies 
 \begin{equation} \label{hbh}
 \qa = \ln \pp1{1 + \f{1}{N}}
  \approx \f{1}{N} .
 \end{equation}
The net force at $t=0$ is thus
 \begin{equation} \label{hbi}
  \Delta f (0) = \Delta f_0 \, N_0 = N .
 \end{equation}

When the temperature grows from zero, $\qa$ is expected to change continuously with $t$, and we may define the low
temperature regime for bosons by $\qa < 1$ so that $\e{-\qa} \approx 1$ along with
$t < 1$. Then we can write for $k>0$,
 \begin{equation}\label{hbj}
  N_k  =  \f{1}{\e{\qa + bk} - 1} = \f{\e{-\qa -bk}}
  {1 - \e{-\qa -bk}} 
  \approx  \e{-bk} + \e{-2bk} + \e{-3bk} + \cdots,
 \end{equation}
where one can expect that, when $n$ is high enough so that $\e{-n\qa}$ ceases to
be near 1, the factor $\e{-nbk}$ appearing in the $n$th term becomes very small.  Consequently,
in this low temperature regime, we have
 \begin{equation} \label{hbk}
  N_0 = N - \sum_{k=1}^\infty N_k \approx N - \e{-b} - 2 \e{-2b}
  - 2 \e{-3b} - \cdots \, .
 \end{equation}
Plugging (\ref{hbj}) and (\ref{hbk}) into (\ref{haf}) and arranging terms according to the powers of $\e{-b}$, we find 
 \begin{equation} \label{hbl}
  \Delta f \approx N - \f{1}{2} \e{-b} - \f{9}{8} \e{-2b} - \f{19}{16} \e{-3b} +
  \cdots \, .
 \end{equation}
We can see that, in the bosonic case, the temperature dependence of $\Delta
f$ begins at the order of $1/N$ with respect to the zero temperature
value.

\section{The medium-temperature regime}
\label{s5}

We have so far gained reasonably good approximations of the net force for high and low-temperature regimes.  In this section we wish to study the temperature regime between the two, hoping to find some 
approximation to interpolate the previous two.  Again, we consider the cases of fermions and bosons, separately.

\subsection{Fermions at medium temperature}

We have seen  for fermions that, at $t \sim 1$, the net force $\Delta f(t)$ is still very close
to $\Delta f(0)$.  For a noticeable departure from $\Delta f(0)$, one
needs to go above $t \sim 1$, presumably to $t \gg 1$ (or $b \ll 1$), in which case
the sum (\ref{hae}) can be approximated by the integral,
 \begin{equation} \label{hbm}
\!\!\! \!\!\! \!\!\! \!\!\! \!\!\!  \!\!\! \!\!\!  \!\!\! \!\!\!  
  N 
  = \sum_{k=0}^\infty
  \f{1}{\e{\qa + bk} - 1} = \frac{1}{\Delta y} \sum_{k=0}^\infty
  \f{\Delta y}{\e{\qa + y_k} + 1}
  \approx \frac{1}{\Delta y} \int_0^\infty \f{\d y}
  {\e{\qa + y} + 1} = \f{1}{b} \ln \pp1{1 + \e{-\qa}} \, ,
 \end{equation} 
where we have used $y_k := bk$, $\Delta y := y_{k+1} - y_k = b$.
This implies
  \begin{equation} \label{hbn}
 \e{-\qa} \approx \e{\f{N}{t}} - 1,
 \end{equation}
which is just (\ref{hak}), showing
that the formula remains valid for medium temperatures as well.

Precise evaluation of $ \e{-\qa} $ is particularly important in this temperature regime, because the final outcome of the net force $\Delta f$ is extremely sensitive to the variation of $ \e{-\qa} $.  For this reason, it is worthwhile to consider an improved
approximation obtained from the fact that, for \lq well-behaved\rq\ functions $g(y)$  in $ [ y_0, \infty ) $ with $\lim_{y \to \infty} g(y) =
0 $, the trapezoid approximation of
integrals yields
 \begin{equation} \label{nbc}
  \sum_{k=0}^{\infty} g(y_k) \approx \frac{g(y_0)}{2} + 
  \frac{1}{\Delta y} \int_{y_0}^{\infty} g(y) \, \d y,
 \end{equation}
which is better than that acting in
(\ref{hbm}) by one order.   As a result, one finds that the first correction term on the rhs of
(\ref{nbc}) improves (\ref{hbn}) to
 \begin{equation} \label{hbo}
  \e{-\qa} \approx \e{\f{N - N_0/2}{t}} - 1 .
 \end{equation}
This formula (\ref{hbo}) reduces to (\ref{hbn}) for high temperatures where
$N_0 \ll 1 \ll N$, and to (\ref{hbc}) for low temperatures where $N_0
\approx 1$ and the second term in (\ref{hbo}) is negligible
compared to the first term. Note that the presence of $N_0 \le 1$ in the improvement is
not insignificant, since at low temperatures
$\e{\f{N - 1/2}{t}}$ can be significantly different from $\e{\f{N}{t}}$ on account of the ratio 
being $\e{\f{1}{2t}} \not \approx 1$. It also suggests that the deviation of $N_0$
from its zero temperature value $1$ may become important for medium
temperatures, too.

\begin{figure}[t]
\vskip 4ex
\centering
\resizebox{.37\textwidth}{!}{\includegraphics
{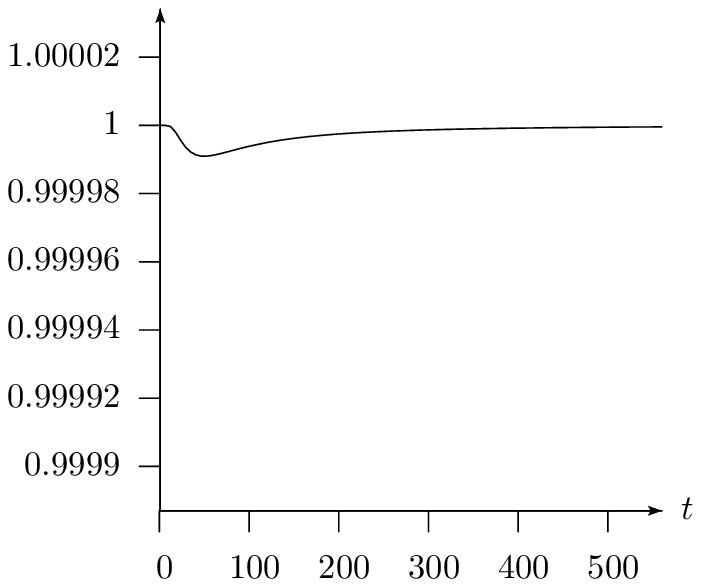}}  \hspace{10mm}
\resizebox{.35\textwidth}{!}{\includegraphics
{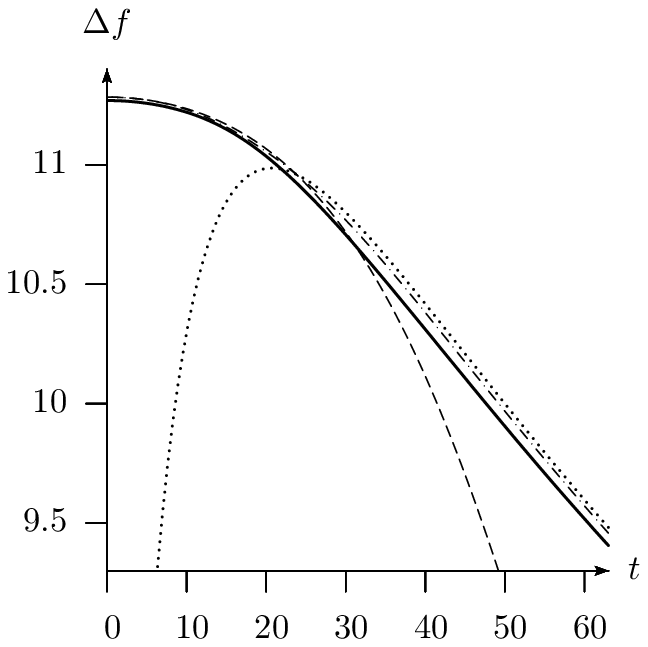}}
\caption{
Fermions, $N = 100$. a) The ratio of the approximate $\e{-\qa}$ provided
by (\ref{hbp}) to the numerically determined $\e{-\qa}$. The
largest relative error is $10^{-5}$ along the whole temperature region
$0 < t < \infty$.
 \,\, b) The net force $\Delta f$. Solid line: the numerical
result. Dashed line: the low+medium-temperature approximation
(\ref{hbx}). Dotted line: the high-temperature approximation
(\ref{hax}). Dash-dotted line: the interpolating curve (\ref{hbz}) with the choices mentioned in the text.
}\label{hbq}
\end{figure}

To acquire a meaningful formula between the low and high-temperature
regimes, we can use (\ref{hbo}) where the presence of $N_0$ is expected to provide
sensitivity to the low temperature regime, with the approximation that $N_0$ is given by  its high temperature value from (\ref{hbn}), $\; N_0
\approx 1 - \e{-\F{N}{t}} \;$, that is,
 \begin{equation} \label{hbp}
  \e{-\qa} \approx 
  \e{(N - \f{1- \e{-\F{N}{t}}}{2} )/t} - 1 .
 \end{equation}
As can be seen in Figure~\ref{hbq}a, this approximation formula for $\e{-\qa}$ holds actually very
well on the whole temperature regime.

Similar improvement for the net force is a bit harder to achieve.  Here, we
will restrict ourselves only to leading-order approximations, and 
proceed as
 \begin{eqnarray} 
&&  \!\!\! \!\!\! \!\!\! \!\!\! \!\!   \Delta f  \approx \sum_{k=0}^\infty
  \f{1}{\s{ \pi \pp1{ k + \F{1}{4} } }} \f{1}{\e{\qa + bk} - 1} =
  \frac{1}{\Delta y} \s{\f{b}{\pi}} \sum_{k=0}^\infty
  \f{\Delta y}{\s{y_k + b/4} \pp1{ \e{\qa + y_k} + 1}} \nonumber \\
&&  \!\!\! \!\!\!  \approx  \frac{1}{\s{\pi b}} \int_0^\infty \f{\d y}
  {\s{y + b/4} \pp1{\e{\qa + y} + 1}},
   \end{eqnarray}
using (\ref{hbs}) again, and considering $b$ small enough. 
Introducing $z := \s{y+b/4}$, we can rewrite it as 
 \begin{equation} \label{hbr}
  \Delta f  \approx 
  \f{2}{\s{\pi b}} \int_{\!\s{b/4}}^\infty \;
  \f{\d z}{\e{\qa - b/4 + z^2} + 1}  
  \approx 
  \frac{2}{\s{\pi b}} \int_0^\infty \f{\d z}{\e{\qa + z^2} + 1}.
 \end{equation}
{}For negative
$\qa$, the resulting integral can be approximated by an asymptotic
series \cite{GS,MS} to obtain,
 \begin{equation} \label{hbv}
  \Delta f  \approx  \frac{2}{\s{\pi b}} \s{-\qa}
  \pp2{ 1 - \f{\pi^2}{24} \f{1}{\qa^2} - \f{7\pi^4}{384} \f{1}{\qa^4}
  + \Ordo{ \f{1}{\qa^6} } }.
 \end{equation}
The divergence of the formula for $\qa \to 0$ may be dealt with by adopting the Pad\'e
form,
 \begin{equation} \label{hby}
  \Delta f  \approx  \frac{2}{\s{\pi b}} \: \f{\s{-\qa}}
  { 1 + \f{\pi^2}{24} \f{1}{\qa^2} + \f{23\pi^4}{1152} \f{1}{\qa^4} }\,,
 \end{equation}
which improves its validity towards $\qa \to 0$,

In the medium-temperature regime, one may expect that the
temperature is low enough for the second term of (\ref{hbo}) to be
omitted, while it is high enough for $N_0/2$ to be omitted.  If this is the case, one can 
simplify (\ref{hbo}) to
$ \e{-\qa} \approx \e{\f{N}{t}}$ or  $-\qa \approx \f{N}{t}$, 
which can be inserted into (\ref{hby}) to obtain
 \begin{equation} \label{hbx}
  \Delta f  \approx  \s{N} \: \f{\frac{2}{\s{\pi}} }
  { 1 + \f{\pi^2}{24} \pp1{\f{t}{N}}^2 +
  \f{23\pi^4}{1152} \pp1{\f{t}{N}}^4 }\,.
 \end{equation}
Note that this formula takes care of the zero temperature value $\Delta f(0)$
correctly, and that, similarly to (\ref{hax}), the ratio $\Delta f / \s{N}$ is
again a function of $t$ and $N$ only through the combination $t/N$.

Plotting the low+medium-temperature curve (\ref{hbx}), and the high
temperature one (\ref{hax}), we find that the two together actually cover
the whole temperature range (see Figure~\ref{hbq}b).  From a technical viewpoint, we may find it convenient to 
introduce an interpolating function for the two curves connecting smoothly.  This can be accomplished by using,  for example, the scheme,
 \begin{equation} \label{hbz}
  \qf_{\hbox{\scriptsize intp}}(\qx) := \f{1 \cdot \qf_1(\qx) +
  \pp1{ \F{\qx}{\qxx}}^\qp \cdot \qf_2(\qx)}{1 +
  \pp1{\F{\qx}{\qxx}}^\qp} ,
 \end{equation}
where $\qf_{\hbox{\scriptsize intp}}(\qx)$ gives the interpolating function of the two curves described by $\qf_1(\qx)$ and $\qf_2(\qx)$.
The interpolating point $\qxx$ may be chosen as the value where
$\qf_1(\qxx) = \qf_2(\qxx)$ holds, which ensures that at $\qxx$ the weights $\f{1}{1 +
\pp1{\F{\qx}{\qxx}}^\qp}$ and $\f{\pp1{\F{\qx}{\qxx}}^\qp}{1 +
\pp1{\F{\qx}{\qxx}}^\qp}$ are equal irrespective of the value of
$\qp$. 

In our case, we have $x = \F{t}{N}$ and choose the function $\Delta f$ in (\ref{hbx}) for $\qf_1$ and 
the function $\Delta f$ in (\ref{hax}) for $\qf_2$.  Then   
the numerically determined value of the interpolating point for  $N=100$ is found to be 
$\qX{\pp1{\F{t}{N}}} = 0.237845
$.
As for $\qp$, we know that
(\ref{hbx}) is precise up to the order of $\pp1{\F{t}{N}}^5$, and if we maintain
this then we find that the
smallest choice is $p = 5$.  Since this choice does not disturb the
high-temperature expansion either, we may propose the resultant interpolating formula
$\qf_{\hbox{\scriptsize intp}}$ in (\ref{hbz}) as the net force covering the whole temperature region.
Figure~\ref{hbq}b shows that this single formula reproduces the precise curve (obtained numerically) very well. 
Note that this interpolating procedure is actually 
independent of $N$, since, for all $N$s, it involves the same functions
(low and high temperature approximants of $\Delta f/\s{N}$) of the identical
variable $t/N$.

\subsection{Bosons at medium temperature}

This time we are again allowed to replace the sums by integrals 
with the approximation (\ref{nbc}).  Since for low temperatures $N_0$ is much larger than
higher $N_k$, we preserve its discrete value and introduce the
continuous variable only above $k = 1$.  This gives
 \begin{eqnarray} \label{hcc}
  N &=& N_0 + \sum_{k=1}^\infty N_k  \approx N_0 + \f{N_1}{2} +
  \frac{1}{\Delta y} \int_{y_1}^\infty \f{\d y}
  {\e{\qa + y} - 1} \nonumber \\
  &= &  \f{1}{\e{\qa} - 1} + \f{\f{1}{2}}{\e{\qa+b} - 1}
  + \f{1}{b} \ln \f{1}{1 - \e{-\qa-b}} \, .
 \end{eqnarray}
 {}For the medium temperature regime, we find it 
reasonable to assume that it is characterized by $\qa \ll 1$ and $t \gg 1$.  Under this, one may simplify the outcome by putting $\qa = \f{b}{4}$ in all terms except
the first in the expansion (\ref{hcc}) (the advantage
of the specific value $\f{b}{4}$ will turn out soon), since by far the first term $N_0 = \f{1}{\e{\qa} - 1}$ is the most sensitive 
term for the change of $\qa$ on the scale of $b$ or below.  In this
approximation, (\ref{hcc}) leads to
 \begin{equation} \label{hce}
  N_0 = \f{1}{\e{\qa} - 1} \approx N - \f{\f{1}{2}}{\e{\f{5b}{4}} - 1}
  - \f{1}{b} \ln \f{1}{1 - \e{-\f{5b}{4}}} \approx N - \f{2}{5} t -
  t \ln \f{4}{5} t \, .
 \end{equation}

The net force can also be evaluated by treating the sum analogously as
 \begin{eqnarray} \label{hcf}
  \Delta f & = & N_0 + \sum_{k=1}^\infty \Delta f_k N_k \approx N_0 +
  \f{\Delta f_1 N_1}{2} + \f{2}{\s{\pi b}} \int_{\!\s{\f{5b}{4}}}^\infty
  \;\f{\d z}{\e{z^2} - 1} , \nonumber \\
  & \approx &  N_0 + \f{1}{2} \f{\f{1}{2}}{\e{\f{5b}{4}} - 1} +
  \f{2}{\s{\pi b}} \int_{\!\s{\f{5b}{4}}}^\infty
  \; \pp2{ \f{1}{z^2} - \f{1}{2} }\d z \, ,
 \end{eqnarray}
 where we have gone through steps similar to 
 (\ref{hbr}) and introduced an asymptotic
expansion for the integrand to improve it in the dominant
region.  Note that the choice $\qa = \f{b}{4}$ has made this integral
$\qa$-independent.

\begin{figure}[t]
\vskip 4ex
\centering
\resizebox{.36\textwidth}{!}{\includegraphics
{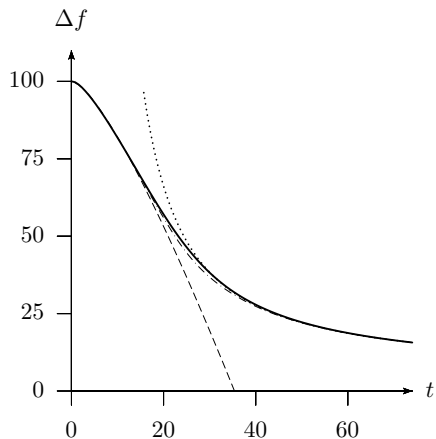}}
\caption{
The net force $\Delta f$ for bosons, $N = 100$. Solid line: the
numerical result. Dashed line: the low+medium-temperature approximation
(\ref{hci}). Dotted line: the high-temperature approximation
(\ref{hat}). Dash-dotted line: the interpolating curve (\ref{hbz}), for
$\qX{t} = 22.925$ and $\qp = 9$.
}\label{hck}
\end{figure}

Actually, the approximation of the integrand with the asymptotic
expansion becomes better if we keep only the part of the integral where the
approximated integrand is positive.  Because of this,  we use
 \begin{equation} \label{hcj}
  \int_{\!\s{\f{5b}{4}}}^{\s{2}} \; \pp2{ \f{1}{z^2} - \f{1}{2} }
  \d z = \s{\f{4}{5} t} - \s{2} + \Ordo{\s{b}} 
 \end{equation}
and (\ref{hce}) to evaluate (\ref{hcf}) and find 
 \begin{equation} \label{hcg}
  \Delta f \approx N + \pp2{ - \f{2}{5} - \ln \f{4}{5} + \f{1}{5} +
  \f{4}{\s{5 \pi}} } t - t \ln t + \s{\f{4}{5} t} - \s{2} .
 \end{equation}
We may further consider some corrections for the expression to render it a little nicer without sacrificing its precision.  
This is done by omitting the last term $\s{2}$ which is negligible compared to the
first term $N$,  which makes the $t = 0$ limiting value the exact value $\Delta f(0) = N$. 
We also replace the
coefficient of $t$ which is numerically $1.032$ by $1$ for simplicity. We then end up with
 \begin{equation} \label{hci}
  \Delta f \approx N + t - t \ln t + \s{\f{4}{5} t} \, .
 \end{equation}
As for lower temperatures the $\qa$-dependence of the terms
$k \ge 1$ is suppressed more, and since
we could reach a formula that is precise
even at $t=0$, we may hope that, in spite of the assumption $\qa \approx
\f{b}{4}$ made above, this formula (\ref{hci}) can be used even for the beginning part of
the net force $\Delta f(t)$ including $t=0$. Figure~\ref{hck} shows that this
is indeed the case.

It is again possible to provide an interpolating formula between
the low-temperature approximation and the high-temperature
one
based on the formula (\ref{hbz}) by choosing the functions (\ref{hci}) and (\ref{hat}) for $\qf_1$ and 
$\qf_2$, respectively.
Note, however, that this time the two curves do not cross each other, and the interpolating point $\qX{t}$ may be chosen,
{\it e.g.}, as the location where the two functions differ the least.
For example, for $N=100$, this yields
$\qX{t} := 22.925$. 
In addition, the value of $\qp$ can be chosen, for
instance, by the requirement that the derivative of the interpolating function 
at $t = \qX{t}$ be the same as that of the straight line
that is the common tangent of the two curves to be connected. 
At $N = 100$, that common tangent straight line touches the low
temperature curve at the point $[7.338, 88.46]$ and the high
temperature curve at $[26.12, 45.66]$ in the coordinate plane of $t$ and
$\Delta f$. The slope of that straight line is $-2.278$, and this is to
be put equal to the derivative of (\ref{hbz}) at $\qxx$, {\it i.e.},
$ 
\f{\d \qf_{\hbox{\scriptsize intp}}}{\d \qx}  (\qxx)
  =   \f{\d \qf_2}{\d \qx} (\qxx) + \f{\qp}{\qxx}
  \f{\qf_2(\qxx) - \qf_1(\qxx)}{4} \, ,
$
where we have used $\f{\d \qf_1}{\d \qx} (\qxx) = \f{\d \qf_2}{\d \qx} (\qxx)$ which comes from the
definition of $\qxx$
being the location where the difference is minimal.   From this we find $\qp = 8.641$. 
One can invent some other criteria as well, but 
our aim here is only to present a formula in which the transition between the two
curves appears as smoothly as possible, yet sharply enough to keep
both curves practically intact on the regions where they are supposed to
be reliable.

Since the accuracy of the interpolation employed above is not very sensitive to the actual value of
$\qp$, we may round it off to the nearest
integer $\qp = 9$, say, for brevity.  Figure~\ref{hck} shows the resultant interpolating formula
(\ref{hbz}) with our choice of functions, which is almost indistinguishable from the
numerical result.  We should, however, keep in mind that for bosons
the interpolation must be done $N$-dependently, because the scaling
behaviour $\f{\Delta f}{\s{N}} = \f{\Delta f}{\s{N}} \pp1{\f{t}{N}}$ seen at
the high-temperature regime does not arise
at the low temperature regime.  It also follows that a successful interpolation at some $N$ does
not necessarily ensure a success for an analogously carried out interpolation at
another $N$.

\section{Conclusion and discussions}
\label{s6}

In this paper, we studied the system of  $N$ particles confined in each of the two half lines separated by a partition wall at the centre of the 
harmonic oscillator potential.  
The partition is assumed to impose a set of distinct  -- the Dirichlet  and the Neumann -- boundary conditions on the left and on the right, respectively.  
Due to the discordance in the energy levels in two sides of the partition, and also to the different distributions of particles on the energy levels at finite temperatures, an induced force emerges on the partition, as we have seen earlier on the partition in the infinite potential well.

We have evaluated the (dimensionless) net force $\Delta f(t)$ that arises on the wall both analytically and numerically as a function of (dimensionless) temperature $t$, and found that it exhibits a number of interesting behaviours characteristic to the harmonic potential.   For instance, it has a non-vanishing limit $\Delta f(0)$ at the zero temperature limit, which is just $N$ for bosons, while it is of the order of $\sqrt{N}$ for fermions.  Note that in the case of the infinite potential well \cite{FTjpa}, the value $\Delta f(0)$ for fermions is proportional to $N^2$.

On the other hand, in the high-temperature regime, the force $\Delta f(t)$ scales linearly in $N$ for both fermions and bosons.   As temperature grows,  it decreases according to $1/\sqrt{t}$ and eventually vanishes in the limit $t \to \infty$.    This is in sharp
contrast to the infinite potential well case where the force $\Delta f(t)$ diverges according to $\sqrt{t}$.

The medium-temperature regime  is somewhat difficult to deal with, but we have succeeded to obtain, after a rather technical argument, an analytic approximation that accounts for the numerical results reasonably well for both fermions and bosons.  Interpolation to the low and high-temperature regimes can also be possible, and we presented a possible formula of the force covering the entire regime of temperature.   Unlike the potential well case,  the net force admits no minimum in the medium-temperature regime.

The characteristic scaling behaviours  in the zero temperature limit $t \to 0$ can be understood heuristically.  
Namely, in the bosonic case, the force
$\Delta f(0)$ comes entirely from the contribution of the ground level where all particles reside, 
and hence it is given by $\Delta f_0$ 
multiplied by the number $N$ of the particles.   In the fermionic case, on the other hand, 
the force $\Delta f(0)$ consists of the contributions of $\Delta f_k  \approx (\pi k)^{-\frac{1}{2}}$ up to the Fermi level, yielding
$\Delta f(0) \propto \sum_{k=1}^N k^{-\frac{1}{2}} \propto N^{\frac{1}{2}}$.   This is to be compared to the infinite potential well case where we have $\Delta f_k \approx k$ and hence $\Delta f(0) \propto \sum_{k=1}^N k\propto N^{2}$.  

In the high-temperature regime, it can also be argued that the steady decrease in the present harmonic potential case, rather than the increase to infinity observed in the potential well case, derives basically from the spectral structure of the harmonic system.   That is,  for higher $n$ the energy level difference between the two half harmonic systems remains constant and does not 
give larger contributions, in contrast to the potential well case where the energy level difference becomes larger and eventually diverges for $n \to \infty$.
In this respect, we recall the upper bound (\ref{ham}) for the force
$\Delta f(t)$ whose decrease for higher $t$ is already expected there as well.
In physical terms, this is also understood from the infinite stretch of the harmonic potential where the
higher energy states can spread more in space.  As a result, 
the energy of the system becomes less sensitive to a shift of the partition at the
centre, resulting in the decrease in the force for high temperatures.  

What can we learn from the results obtained here for the harmonic system when combined with those obtained previously for the infinite potential well system?   Suppose that a partition wall can actually be manufactured with the distinct set of boundary conditions assumed in this paper, and further that the induced force on the partition wall can be measured with sufficient accuracy.  Then, one can estimate the profile of the potential in the neighbourhood of the partition wall by looking at the low temperature behaviour of $\Delta f$, since the net force is sensitive to the spectral structure up to the Fermi level for fermions when the temperature is low.   From the high temperature behaviour of $\Delta f$, one can also obtain a crude picture of how the potential stretches in space further away from the wall.  In addition, the characteristic scaling in the number $N$ of particles in the zero temperature limit will reveal the statistics of the particles contained around the wall.    In short, such a nontrivial partition wall may be quite useful in probing the profile of the potential as well as the statistics of the particles involved. The present probe is still primitive to use for generic potentials, but it can be improved if we learn further the behaviours of the net force for other types of potentials in the high and the low temperature regimes along with the scaling property in the particle number.

\ack{
This work has been supported in part by
the Grant-in-Aid for Scientific Research (C), No.~20540391-H20, MEXT, Japan.
}


\end{document}